\documentclass[%
 reprint, superscriptaddress,
 amsmath,amssymb,
 aps,
a4paper,
accepted=2023-09-27
]{quantumarticle}
\pdfoutput=1

\usepackage[english]{babel}
\usepackage{blindtext}
\usepackage{physics, braket, amsmath, xfrac, hyperref, siunitx, dsfont}
\usepackage{graphicx}% Include figure files
\usepackage{dcolumn}% Align table columns on decimal point
\usepackage{bm}% bold math
\usepackage{framed}
\usepackage{multirow}
\usepackage{natbib}
\usepackage[utf8]{inputenc}
\usepackage{pgfplots}
\DeclareUnicodeCharacter{2212}{−}
\usepgfplotslibrary{groupplots,dateplot}
\usetikzlibrary{patterns,shapes.arrows}
\pgfplotsset{compat=newest}

\newcommand{\valerio}[1]{\textcolor{black}{#1}}

\newcommand{\revun}[1]{\textcolor{black}{#1}}
\newcommand{\revdeux}[1]{\textcolor{black}{#1}}
\newcommand{\revtrois}[1]{\textcolor{black}{#1}}
\newcommand{\revother}[1]{\textcolor{black}{#1}}
\newcommand{\final}[1]{\textcolor{black}{#1}}

\hypersetup{
    colorlinks,
    citecolor=blue,
    filecolor=blue,
    linkcolor=blue,
    urlcolor=blue
}

\begin{document}

\title{Neural Network Approach to the Simulation of Entangled States with One Bit of Communication}% Force line breaks with \\
% \thanks{A footnote to the article title}%

\author{Peter Sidajaya}
\email{peter.sidajaya@u.nus.edu}
\affiliation{Centre for Quantum Technologies, National University of Singapore, 3 Science Drive 2, Singapore 117543}
\orcid{0000-0002-5003-6242}

\author{Aloysius Dewen Lim}
\affiliation{Department of Physics, National University of Singapore, 2 Science Drive 3, Singapore 117542}

\author{Baichu Yu}
\affiliation{Centre for Quantum Technologies, National University of Singapore, 3 Science Drive 2, Singapore 117543}
\affiliation{Shenzhen Institute for Quantum Science and Engineering, 
Southern University of Science and Technology, Nanshan District, Shenzhen, 518055, China}
\affiliation{International Quantum Academy (SIQA), Shenzhen 518048, China}
\orcid{0000-0001-8354-8343}

\author{Valerio Scarani}
\affiliation{Centre for Quantum Technologies, National University of Singapore, 3 Science Drive 2, Singapore 117543}
\affiliation{Department of Physics, National University of Singapore, 2 Science Drive 3, Singapore 117542}
\orcid{0000-0001-5594-5616}

\date{30 September 2023}% It is always \today, today,
             %  but any date may be explicitly specified

\begin{abstract}
% From Bell’s theorem we know that local hidden variables could not simulate the behaviour of an entangled state measured using projective measurements. Surprisingly, by adding one bit of communication between the parties, it is known that it is enough to simulate the behaviour of Bell state. Unfortunately, extending this to the case for a non-maximally entangled two-qubit state and states of higher dimension proved to be difficult. Here, we tried to use an Artificial Neural Network (ANN) constrained by locality and supplemented by one bit of communication to explore these open questions. Our results suggest that it might be possible to simulate the behaviour of a non-maximally entangled two-qubit states using only one bit.

Bell's theorem states that Local Hidden Variables (LHVs) cannot fully explain the statistics of measurements on some entangled quantum states. It is natural to ask how much supplementary classical communication would be needed to simulate them. We study two long-standing open questions in this field with neural network simulations and other tools. First, we present evidence that all projective measurements on partially entangled pure two-qubit states require only one bit of communication. We quantify the statistical distance between the exact quantum behaviour and the product of the trained network, or of a semianalytical model inspired by it. Second, while it is known on general grounds (and obvious) that one bit of communication cannot eventually reproduce all bipartite quantum correlation, explicit examples have proved evasive. Our search failed to find one for several bipartite Bell scenarios with up to 5 inputs and 4 outputs, highlighting the power of one bit of communication in reproducing quantum correlations.
\end{abstract}

\maketitle

\section{Introduction}

% Since Bell's theorem, we have known that Local Hidden Variables (LHVs) are inadequate for fully describing the behaviours of entangled quantum states \cite{bell1964einstein}. However, merely violating a Bell inequality does not provide us with an intuition for the power of an entangled state. Therefore, some have posed the question of \textit{"How much additional resources, on top of the LHVs, are needed to simulate quantum behaviour with local resources?"} . While some works have been done with nonlocal boxes as supplementary resources \cite{popescu1994quantum,brunner2005entanglement,brunner2008simulation}, we are interested in the question of how much \textit{classical communication} is needed to simulate the behaviour of a given state \cite{steiner2000towards, csirik2002cost, toner2003communication}. This approach is more intuitive, since, unlike nonlocal boxes, classical communication is a physical resource that we commonly use. 

Quantum Mechanics is famous for having randomness inherent in its prediction. Einstein, Podolski and Rosen argued that this makes quantum mechanics incomplete, and suggested the existence of underlying Local Hidden Variables (LHV) \cite{einstein1935can}. While this view was disproved by Bell's theorem \cite{bell1964einstein, scarani2019bell}, it has nevertheless proved fruitful to approach quantum correlations, without committing to an ontology of the quantum world, by asking \textit{which resources would one use to simulate them}. Though insufficient, LHV provide an intuitive starting point -- then, the question becomes: \textit{which additional resources, on top of the LHV, are needed to simulate quantum correlations?}. Some works have considered nonlocal boxes as supplementary resources \cite{popescu1994quantum,brunner2005entanglement,brunner2008simulation}: while appealing for their intrinsic no-signaling feature, these hypothetical resources are as counterintuitive as entanglement itself, if not more. Classical communication, on the other hand, is a resource that we use on a daily basis and of which therefore we have developed an intuitive understanding. Because we are thinking in terms of simulations and not of ontology, we are not impaired by the very problematic fact that communication should be instantaneous if taken as the real underlying physical mechanism.

Therefore, we are interested in the question of how much classical communication must supplement LHV to simulate the behaviour of a quantum state. For the maximally entangled state of two qubits, after some partial results \cite{brassard1999cost, steiner2000towards, csirik2002cost}, Toner and Bacon provided a definitive solution by describing a protocol that simulates the statistics of all projective measurements using only one bit of communication, which we refer to as \textit{LHV+1} \cite{toner2003communication}. Subsequently, Degorre and coworkers used a different approach and found another protocol which also requires only one bit of communication \cite{degorre2005simulating}. The case of non-maximally entangled pure states proved harder. By invoking the Toner-Bacon model, two bits of communication are certainly sufficient \cite{toner2003communication} \revtrois{ and it was recently proved that two bits are also enough for POVM measurements \cite{renner2023classical}. Meanwhile, Brunner and coworkers proved that one PR-box is not enough for projective measurements \cite{brunner2005entanglement}.} But the simulation of those states in LHV+1 remained open. Only recently, Renner and Quintino reported an LHV+1 protocol that simulates exactly weakly entangled pure states \cite{renner2022minimal}. Our neural network will provide evidence that projective measurements on all two-qubit states can be very closely approximated in LHV+1.

The LHV+1 problem could, in principle, be approached systematically, since the behaviours that can be obtained with those resources are contained in a \textit{polytope}. However, the size of this polytope grows very quickly with the number of inputs and outputs: as of today, after some initial works \cite{bacon2003bell,maxwell2014bell}, the largest LHV+1 polytope to be completely characterized has three measurements per party and binary outcomes; and no quantum violation is found \cite{zambrini2019bell}. \revtrois{Addressing the problem for higher-dimensional systems has also been challenging. Some results has been found for the average amount of communication \cite{massar2001classical, degorre2007classical, brassard2019remote}. On the other hand, for the minimum amount, Brassard et al. showed that for $n$ pairs of Bell states, the amount of communication necessary must grow as $2^n$ \cite{brassard1999cost}. This clearly shows that one bit \textit{must} fail to suffice at some point, but \final{the question of which states can or cannot be simulated by it remains open.} Finally, in the finite output scenario, Vértesi and Bene showed that a pair of maximally entangled four-dimensional quantum systems cannot be simulated with only one bit of communication by presenting a scenario involving an infinite number of measurements \cite{PhysRevA.80.062316}. Here, we will try to answer whether there is a finite scenario where one bit of communication fails to simulate a quantum correlation.}

In recent years, there have also been increasing attempts to study quantum correlations with machine learning. Many of them reveal the great potential neural network has in tackling the complexities in detecting nonlocality and entanglement \cite{deng2018machine, ma2018transforming, canabarro2019machine, krivachy2020neural, harney2021mixed, girardin2022building}. The choice of tackling the LHV+1 problem with machine learning is prompted by the fact that there is no compact parametrisation of LHVs, nor of the dependence of the bit of communication from the parameters of the problem. Thus, we are looking for a solution to a problem, whose variables are themselves poorly specified. Moreover, similar to an LHV model, everything inside a neural network has definite values. Thus, it seems natural to devise a machine learning tool, specifically an artificial neural network (ANN), to act as an LHV model.

% \valerio{Do you agree with this?}

This work is separated into two sections. In Section \ref{sec:two}, we study the simulability of the correlations of entangled state with classical resource and one bit of communication using a neural network. We also present a semianalytical protocol which \textit{approximates} the behaviour of partially entangled two-qubit states with one bit of communication, and we also study the errors of our protocol. In Section \ref{sec:polytope}, we also try to find a quantum behaviour in dimensions higher than two qubits that could not be simulated by a single bit of communication.

\section{Simulating Two-qubit Entangled States using Machine Learning}
\label{sec:two}

\begin{figure*}
    \centering
    \includegraphics[width=\textwidth]{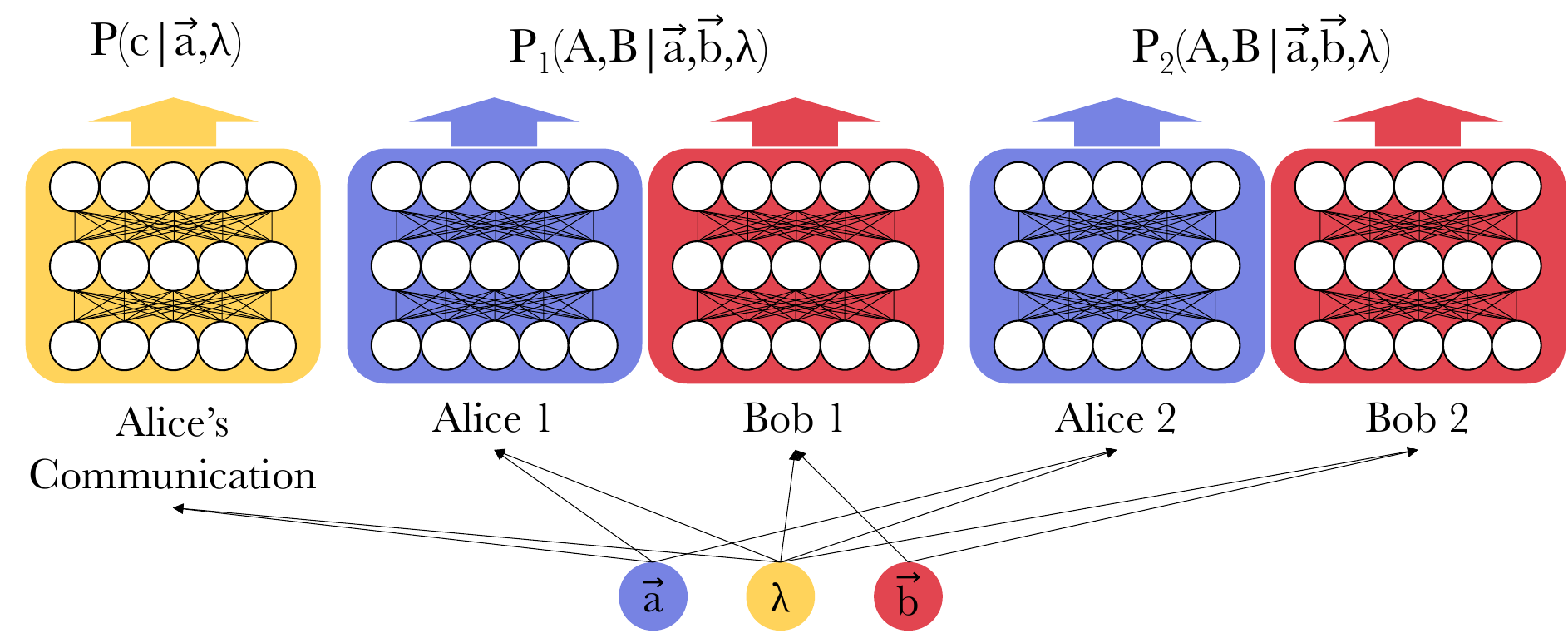}
    \caption{The architecture of the Artificial Neural Network (ANN). The model consists of two local distributions and a communication network. In each distribution, the two parties are constrained by locality by routing the input accordingly. The communication network outputs a value between 0 and 1, and represents the probability of Alice sending a certain bit to Bob. The output for a particular round is then simply the convex combination of the two local distributions.}
    \label{fig:ANN_diagram}
\end{figure*}

\subsection{Using Neural Network to generate protocols}

Inspired by the use of a neural network as an oracle of locality \cite{krivachy2020neural}, we approached the problem using an artificial neural network. The network takes in measurement settings $\Vec{a}$ and $\Vec{b}$ as an input and outputs an LHV+1 bit probability distribution, enforced by an architecture that forces the suitable locality constraints, which we will discuss below. The output distribution is then compared against the target distribution using a suitable error function, such as the Kullback-Leibler divergence.

The Local Hidden Variables (LHV) are described by a random variable $\lambda$ shared among both parties. \revother{For a finite communication model to work, $\lambda$ needs to be a random variable of infinite length \cite{massar2001classical}. Besides that,} $\lambda$ can be of any form. The analytical model of Toner and Bacon uses a pair of uniformly distributed Bloch vectors, that of Renner and Quintino \cite{renner2022minimal} uses a biased distribution on the Bloch sphere. The neural network of \cite{krivachy2020neural} \valerio{simulates a model where the LHV is a single real number, distributed normally or uniformly}. In theory, the choice is ultimately redundant because the different LHV models can be made equivalent by some transformation. However, the neural network will perform differently since it can only process a certain amount of complexity in the model. From trial and error, we settled on Toner and Bacon's \textit{uniformly distributed vector pair} as the LHV model in our neural network.

% \valerio{where does it matter that it's a vector? And are you not confusing the vector with its distribution?} 

A probability distribution $P(A,B)$ is \textit{local} if it can be written as
\begin{equation}
    \begin{aligned}
    P_L(A,B\mid \Vec{a},\Vec{b})
    &= \int P(A\mid \Vec{a}, \lambda) \; P(B\mid \Vec{b}, \lambda) \; d\lambda.
    \end{aligned}
    \label{eq:LHV}
\end{equation}
The network approximates a local distribution by the Monte Carlo method as
\begin{equation}
    \begin{aligned}
    P_L(A,B\mid \Vec{a},\Vec{b})
    &= \frac{1}{N}\sum_{i=1}^N P(A\mid \Vec{a}, \lambda_i) \; P(B\mid \Vec{b}, \lambda_i),
    \end{aligned}
    \label{eq:LHV_approx}
\end{equation}
where $N$ is a sufficiently large number ($\geq 1000$). In the network, Alice and Bob are represented as a series of hidden layers. Each of the parties takes in their inputs according to the locality constraint and outputs their own local probability distribution. The activation functions used in the hidden layers are the standard functions, such as the rectified linear unit (ReLU) and the softmax function used to normalise the probabilities. The forward propagation is done $N$ times using varying values of $\lambda_i$ sampled from the chosen probability distribution. Thereafter we take the average of the probabilities over $N$ to get the probability distribution as expressed in equation (\ref{eq:LHV_approx}).

To move from LHV to LHV+1, we notice that sending one bit of communication is equivalent to giving Alice the power of making the decision to choose between one out of two local strategies. The recipe looks as follows:
\begin{itemize}
\item Alice and Bob pre-agreed on \textit{two} local strategies $P_{L,1}$ and $P_{L,2}$, as well as on the $\lambda$ to be used in each round. It seems to us that all previous works in LHV+1 assumed $P_1(A\mid \Vec{a}, \lambda)=P_2(A\mid \Vec{a}, \lambda)$, but of course there is no need to impose such a constraint.
\item Upon receiving her input $\Vec{a}$, Alice decides which of the two strategies should be used for that round, taking also $\lambda$ into account. Some of the previous LHV+1 models used a deterministic decision model, but there is no reason to impose that: Alice's decision could be stochastic. She informs Bob of her choice with one bit of communication $c$, and Bob consequently keeps his outcome for the chosen strategy.
\end{itemize}
Thus, given a randomly sampled LHV $\lambda_i$, the LHV+1 model is described by
\begin{widetext}
\begin{equation}
    \begin{aligned}
      P(A,B\mid \Vec{a}, \Vec{b}, \lambda_i)
      &= P(c=+1\mid \Vec{a}, \lambda_i) P_{L,1}(A,B\mid \Vec{a},\Vec{b}, \lambda_i) + P(c=-1\mid \Vec{a}, \lambda_i) P_{L,2}(A,B\mid \Vec{a},\Vec{b}, \lambda_i) \\
      &=P(c=+1\mid \Vec{a}, \lambda_i) P_1(A\mid \Vec{a}, \lambda_i)P_1(B\mid \Vec{b}, \lambda_i) + P(c=-1\mid \Vec{a}, \lambda_i) P_2(A\mid \Vec{a}, \lambda_i)P_2(B\mid \Vec{b}, \lambda_i).
      \label{eq:comm}
    \end{aligned}
\end{equation}
\end{widetext} where we labeled $c=+1$ (respectively $c=-1$) the value of the bit of communication when Alice decides for strategy 1 (resp. 2).

Now the complete model consists of two local networks and one communication network. The communication network consists of a series of layers whose inputs are the same as Alice's and outputs a number between 0 and 1 by using a sigmoid activation function, representing $P(c \mid \Vec{a}, \lambda_i)$, which then is used to make a convex mixture of the two local strategies, for the particular inputs and LHV. The final network architecture, then, can be seen in Fig. \ref{fig:ANN_diagram}.

This approach of using a neural network to generate local strategies was originally used in a network setting \cite{krivachy2020neural}. In that work, the network was used to verify nonlocality by looking for transitions in the behaviours of distributions when mixed with noise. When a state is mixed with noise, it lies within a local set, up to a certain noise threshold; reducing the amount of noise in the state allows for the identification of sharp transitions in the network's error, indicating when the state exits the local set. Here, instead of such an oracle, we will use the network to generate a protocol to simulate the quantum state by analysing its outputs.

\subsection{Simulating Two-qubit States}

For a two-qubit scenario, the joint measurements can be defined by two vectors in the Bloch sphere, i.e. $\Vec{a}, \Vec{b} \in S^2$. \revdeux{Here, we are only considering projective measurements with binary outputs. Thus, the projectors are defined by
\begin{align*}
    \Pi^{\Vec{a}}_{A} = \frac{\mathds{I} + (-1)^A\,\Vec{a}\cdot\Vec{\sigma}}{2},
\end{align*}
\revdeux{and similarly for Bob. The behaviour is the set}
\begin{widetext}
\begin{equation}
    \mathcal{P}(\rho) = \left\{ \left\{ P_\rho(A,B\mid \Vec{a},\Vec{b}) = \Tr(\Pi^{\Vec{a}}_{A} \otimes \Pi^{\Vec{b}}_{B} \, \rho) \; \Big| \; \forall A,B \in \{+1,-1\} \right\}
    \; \Big| \; \Vec{a}, \Vec{b} \in S^2 \right\},
\end{equation}
\end{widetext}
i.e., the set of correlations for all possible measurement directions.}

\subsubsection{Maximally Entangled State}\label{subsubsec:max-entangled}

The maximally entangled state case has been solved analytically by Toner and Bacon \cite{toner2003communication}. Thus, we used this state as a test bed for our machine learning approach by training the machine to simulate the distribution of the maximally entangled state $\ket{\Psi^-}$.

\revun{A snapshot of the behaviour of the trained model can be seen in Fig. \ref{fig:pi-4} in Appendix \ref{app:figures}. These figures, along with others which can be generated with the code, are very similar to the figures in the paper of Toner and Bacon, with theirs being plotted on a sphere and ours being a projection. The major difference is, in our plots, Alice's output is different depending on the bit of communication. By comparing our guesses and the plots generated by the neural network, we deduced that the behaviours of the parties in the neural network are: }

\begin{framed}
Maximally entangled state protocol:
\begin{enumerate}
    \item Alice sends to Bob
    $$c=\text{sgn} {(\Vec{a} \cdot \Vec{\lambda}_1)} \text{sgn} {(\Vec{a} \cdot \Vec{\lambda}_2)},$$
    \item Alice outputs
    $$A=-\text{sgn} (\Vec{a} \cdot (\Vec{\lambda}_1 + c \Vec{\lambda}_2)),$$
    \item Bob outputs
    $$B=\text{sgn} (\Vec{b} \cdot (\Vec{\lambda}_1 + c \Vec{\lambda}_2)),$$
\end{enumerate}
\revtrois{where $\text{sgn}(x) = \frac{x}{|x|}$ is the sign function.}
\end{framed}

The protocol bears much resemblance to Toner and Bacon's original protocol, with the only difference being the output of Alice, which are simply $-\text{sgn}( \Vec{a} \cdot \Vec{\lambda}_1)$ in the original protocol.

\revdeux{We can further check that this is model reproduces the correct correlation by checking the expected marginals of the maximally entangled state:
\begin{align}
    \langle A \rangle &= \iint \; A \; d\Vec{\lambda}_1 d\Vec{\lambda}_2 = 0, \nonumber \\
    \langle B \rangle &= \iint \; B \; d\Vec{\lambda}_1 d\Vec{\lambda}_2 = 0, \nonumber \\
    \langle AB \rangle &= \iint \; AB \; d\Vec{\lambda}_1 d\Vec{\lambda}_2 = - \Vec{a} \cdot \Vec{b}.
\end{align}
While the first two equations can be proven analytically to be correct for our functions using the same method used by Toner and Bacon, proving the third equality is more difficult. However, numerical integration shows that the third equation also holds up to arbitrary numerical precision, depending on the accuracy set for the numerical integration, for uniformly sampled combinations of $\Vec{a}$ and $\Vec{b}$.}

\subsubsection{Non-maximally Entangled States}

\begin{figure}[h!]
    \centering
    \includegraphics[width=\linewidth]{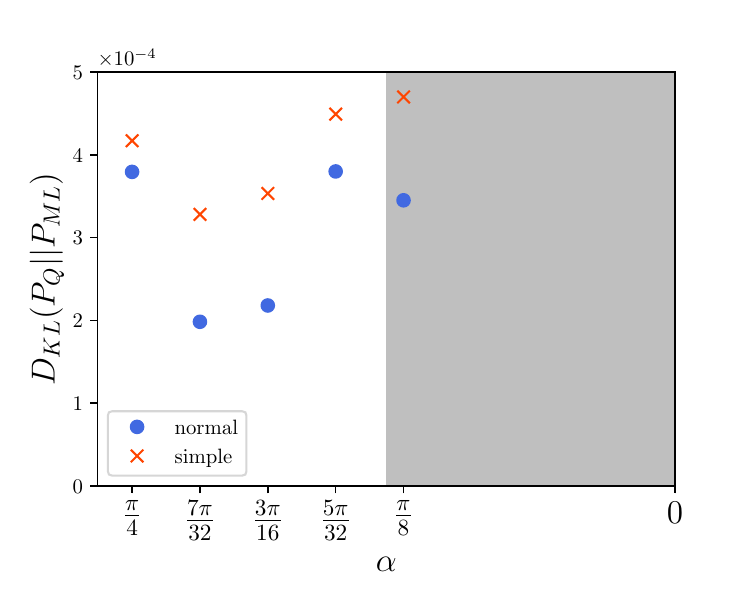}
    \caption{The relative error between the neural network models' behaviours and the quantum behaviours. The blue dots are the original model described, while the red crosses are the simplified model described in the text. The grey shaded region is the region in which an LHV+1 model is known \cite{renner2022minimal}.}
    \label{fig:graph}
\end{figure}

We now apply the same method to the non-maximally entangled two-qubit states. Without loss of generality, any pure two-qubit states can be written in the form of
$$
\ket{\psi(\alpha)} = \cos(\alpha) \ket{01} - \sin(\alpha) \ket{10}, \quad \alpha \in \left[0, \frac{\pi}{4}\right]
$$
using a suitable choice of bases. The state is maximally entangled when $\alpha = \frac{\pi}{4}$ and separable when $\alpha = 0$. We trained the network to simulate the distribution of $\ket{\psi(\alpha)}$ with $\alpha \in \left[0, \frac{\pi}{4}\right]$. A selection of the resulting protocol is shown in Fig. \ref{fig:5pi-32}.

The errors of the models for these states are \revdeux{similar to the one} for the maximally entangled state (see Fig. \ref{fig:graph}). This does not necessarily mean that they successfully simulate the states exactly, instead of simply approximating them. If the behaviour were actually nonlocal, we should expect a transition in the error when we mix the state with noise, signifying the exit of the state from the local+1 bit set. However, we observe no clear transition occurring when noise is added to the state, only a shallow gradient, suggesting that it is still inside the local+1 bit set. While encouraging, this does not constitute a proof, and we would still need to write an analytical protocol. Unlike for the case of maximally entangled states, the models we obtained for the non-maximally entangled states are more complex and our attempt to infer a protocol begins by looking at figures similar to Fig. \ref{fig:5pi-32} \revdeux{in Appendix \ref{app:figures}.}

We start from the parties' outputs:
\begin{framed}
The outputs of Alice are of the form of
\revtrois{
$$P(A_1\mid\Vec{a},\Vec{\lambda}_1,\Vec{\lambda}_2)=\frac{1}{2} (1-A_1\,\text{sgn}(\Vec{a} \cdot \Vec{\lambda}_{a1} + b_{a1})),$$}
where $\Vec{\lambda}_{a1} = u_{a1} \Vec{\lambda}_1 + \Vec{\lambda_2} + v_{a1} \Vec{z}$ decides the hemisphere direction and $b_{a1}=w_{a1}+x_{a1}\Vec{\lambda}_1\cdot \Vec{z}+y_{a1}\Vec{\lambda}_2\cdot \Vec{z}$ decides the size of the hemisphere. Similarly,
\revtrois{
$$P(A_2\mid\Vec{a},\Vec{\lambda}_1,\Vec{\lambda}_2)=\frac{1}{2} (1+A_2\,\text{sgn}(\Vec{a} \cdot \Vec{\lambda}_{a2} + b_{a2})),$$
$$P(B_1\mid\Vec{b},\Vec{\lambda}_1,\Vec{\lambda}_2)=\frac{1}{2} (1+B_1\,\text{sgn}(\Vec{b} \cdot \Vec{\lambda}_{b1} + b_{b1})),$$
$$P(B_2\mid\Vec{b},\Vec{\lambda}_1,\Vec{\lambda}_2)=\frac{1}{2} (1-B_2\,\text{sgn}(\Vec{b} \cdot \Vec{\lambda}_{b2} + b_{b2})).$$}
Using numerical algorithms, we can approximately obtain the relevant coefficients, laid out in \revtrois{Table \ref{tab:party_coefficients_1} in} Appendix \ref{app:tables} for the different states. 
\end{framed}

\revtrois{This guess comes from intuition. There are three points to note. First, Fig. \ref{fig:5pi-32} shows that the outputs of the parties remain hemispherical, meaning that it is still generated from a $\text{sgn}$ function. Thus, we obtain this expression by first fitting the normal direction of the plane in the hemispherical model to the neural network output. By looking at the fitted hemispherical model, one can see that the general movement of the normal direction is a weighted sum of $\Vec{\lambda}_1$ and \final{$\Vec{\lambda}_2$}, thus we have the parameter $u$. Second, the expected values $\langle A \rangle, \langle B \rangle, \langle AB \rangle$ for the non-maximally entangled states are biased in the $\hat{z}$ direction. More specifically, they are
\begin{align}
    \langle A \rangle &= \cos(2\alpha) a_z \nonumber \\
    \langle B \rangle &= -\cos(2\alpha)b_z \nonumber \\
    \langle AB \rangle &= -a_z b_z - \sin(2\alpha) (a_x b_x + a_y b_y).
\end{align}
Thus, we add an arbitrary parameter that allows biases in the form of the parameter $v$. Third, since $\langle A \rangle$ and $\langle B \rangle$ are nonzero, the hemispheres cannot divide the sphere perfectly in two and we must give a bias inside the sign function in the form of $b$, which then gives the parameters $w$, $x$, and $y$.}

\revtrois{Note that since the parameters come from what essentially is a regression, not all of the parameters are needed. In fact, by looking at Table \ref{tab:party_coefficients_1}, many of these parameters are basically zero for many of the states and parties. Moreover, many of the parameters ($b$ and $v$) should reduce to zero for the maximally entangled state in order to obtain the Toner and Bacon model. The fact that $v$ is quite far from zero for the maximally entangled state shows that much of the errors in fitting this model might come from this parameter.}

\revtrois{So far the expression is still integrable. However, things start to become complicated when we move to the bit of communication.} To start, notice  (Fig.~\ref{fig:5pi-32}) that the neural network simulations converge to a solution, in which the bit of communication is not deterministic for some inputs \footnote{This stochasticity could be removed by encoding the randomness in an additional LHV. For example, add an uniform LHV $\lambda_U \sim U(0,1)$. Then, the current strategy could be made deterministic by taking the output of the current communication function as a cutoff. I.e, send +1 if $P(c=+1) \geq \lambda_U$. Thus, it is now deterministic for a given $(\Vec{\lambda}_1,\Vec{\lambda}_2,\lambda_U)$.}. However, guessing such a stochastic communication function proved impossible for us: hence, we tried to force a simpler deterministic model by reducing the communication part of the neural network to a single hidden layer. Albeit having a worse error (see Fig.~\ref{fig:graph}), we managed to obtain a close expression of the output of this new, simplified, network. Its output can be seen in Fig.~\ref{fig:5pi-32-simple} in Appendix \ref{app:figures}.

\begin{framed}
The (simplified) bit of communication is given by
$$
\final{P(c\mid\Vec{a})=\frac{1}{2} (1-c\cdot\text{clip}(f_c,-1,1))},
$$
where
\begin{align*}
    f_c&=\Theta(\Vec{a}\cdot\Vec{\lambda}_1+b_c)\Theta(\Vec{a}\cdot\Vec{\lambda}_2+b_c) \\
    & +\Theta(-\Vec{a}\cdot\Vec{\lambda}_1+b_c)\Theta(-\Vec{a}\cdot\Vec{\lambda}_2+b_c) \\
    & -\Theta(-\Vec{a}\cdot\Vec{\lambda}_1-b_c)\Theta(\Vec{a}\cdot\Vec{\lambda}_2-b_c) \\
    & - \Theta(\Vec{a}\cdot\Vec{\lambda}_1-b_c)\Theta(-\Vec{a}\cdot\Vec{\lambda}_2-b_c),
\end{align*}
with $b_c = u_c + v_c(\Vec{\lambda}_2\cdot \Vec{z})(1-\Vec{\lambda}_1\cdot \Vec{z})$ and the clip function is defined as
$$
\text{clip}(x,a,b) = 
\begin{cases}
    a \quad &\text{if } x < a \\
    b \quad &\text{if } x > b \\
    x \quad &\text{otherwise}
\end{cases},
$$
\revtrois{and $\Theta(x)=\frac{1}{2}(1+\text{sgn}(x))$ is the Heaviside step function.} Again, the relevant coefficients obtained using numerical methods are listed in \revtrois{Table \ref{tab:party_coefficients_1}} in Appendix \ref{app:tables}.
\end{framed}

\revtrois{The intuition behind this choice of seemingly random function comes from the fact that it must reduces to the Toner and Bacon model for the maximally entangled state. When we look at Fig.~\ref{fig:5pi-32-simple} in Appendix \ref{app:figures}, we can see that the form is similar to the usual product of two $\text{sgn}$ functions in the Toner and Bacon model. However, in the original Toner and Bacon model, there is always \final{a pair of antipodal points} bordered by two +1 regions on opposite sides and two -1 regions on opposite sides, forming a quadripoint. These points corresponds to the unique vectors $\pm\Vec{n}$ which are orthogonal to both $\Vec{\lambda}_1$ and $\Vec{\lambda}_2$. \final{This pair of quadripoints cease to exists in the non-maximally entangled states models}, thus there is an 'overflow' or an imbalance in the +1 and -1 regions. In order to mimic this, we first notice that
$$
\text{sgn(x)} = \Theta(x) - \Theta(-x).
$$
Next, we split the two $\text{sgn}$ functions into \final{these} two terms and multiply the two of them, obtaining four terms in total. Then, we add a bias term $b_c$ in each of the terms to mimic the imbalance or the 'overflow' in the different regions. In general, we can set the eight different biases in each \final{Heaviside step function} to be independent of each other, but we decided against \final{that} to prevent overfitting an already complex model so as to make it unreadable. We have decided to place the biases in such configuration to respect the antipodal symmetry of the figures (the transformation $\Vec{a} \rightarrow -\Vec{a}$ will interchange the first and second term and similarly for the third and fourth term) and to oppose the positive terms symmetrically with the negative terms. The final form of the bias itself is found through regression and guessing. Finally, we normalise the function back with the $\text{clip}$ function.}

\revtrois{We call this set of functions which we guess to be the functions of the neural network the \textit{semianalytical} model. Since the results we have presented are numerical in nature, this protocol is not an exact protocol, but simply an approximation. In theory, we could \textit{analytically} integrate these functions and try to match the numerical parameters to the expectation values of the quantum behaviour. However, the communication function is simply too complex for us to perform this integration. Thus, in the next section, we will benchmark the performance of this model along with the neural network with more statistical measures to get an intuition of the 'closeness' of these approximations.}

\revun{It is worth noting that the model suggested by our neural network is very different from that of Renner and Quintino \cite{renner2022minimal}. Notably, in their model, one of the LHV is distributed with a bias; while we have set in our code that the LHV are two uniformly distributed Bloch vector. The fact that there exist different LHV+1 models is not surprising: already for the simulation of the maximally entangled state, the model of Degorre et al.~\cite{degorre2005simulating} differs from that of Toner-Bacon \cite{toner2003communication}.}

\subsection{Statistical Analysis of the Simulations}

After presenting our protocols, we can now consider the performance of our protocols, both the neural network protocol itself and the semianalytical protocol we distilled from it. These LHV+1 protocols are not exact protocols, but approximations, and we can describe their closeness to the quantum behaviour by providing statistical error values. To get a better intuition on the error values, let us consider a hypothesis testing scenario \cite{kullback1997information}. Suppose that we have an unknown sample of length $n$ generated by the same measurement done to $n$ identical systems. Suppose also that we know that the systems are all actual quantum systems ($P_Q$), or our LHV+1 models ($P_{LHV+1}$), but we do not know which. Let us take $P_{LHV+1}$ as the null hypothesis. Let $a$ be the Type I error (mistakenly rejecting a true null hypothesis). In our case, a Type I error would correspond to our machine learning model successfully spoofing as a quantum system. For any decision-making procedure, the probability of a Type I error is lower bounded by $$a \geq e^{-nD_{KL}(P_Q||P_{LHV+1})}.$$ Thus, in order to have 95\% confidence in rejecting a sample from the LHV+1 model, we would need a sample size of
$$
n_{95\%} \geq -\frac{\ln{0.05}}{D_{KL}(P_Q||P_{LHV+1})}.
$$

\begin{figure}[h]
    \centering
    \includegraphics[width=\linewidth]{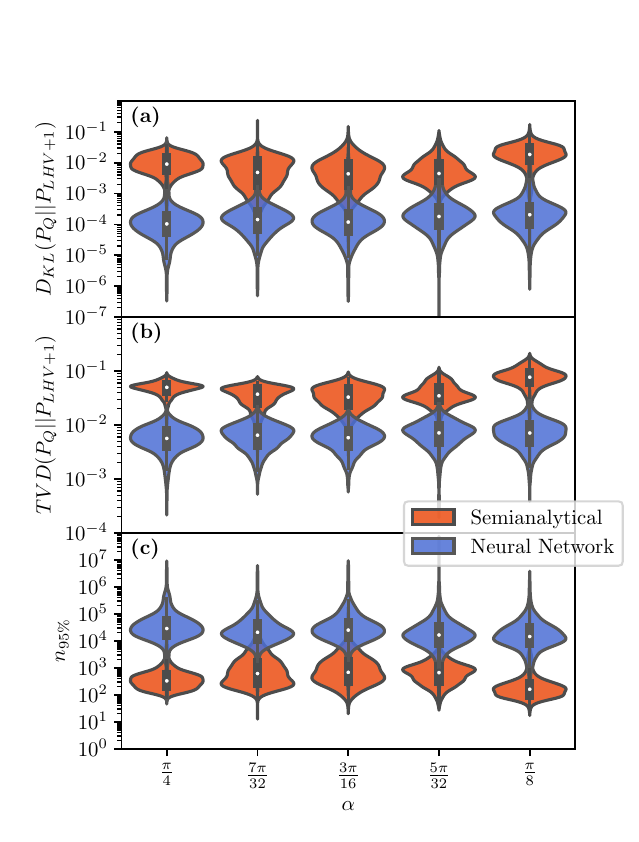}
    \caption{Violin plots for the neural network (blue) and the semianalytical protocol we presented (red) describing the following values: \textbf{(a)} The Kullback-Leibler divergence between our protocols and the quantum behaviours. \textbf{(b)} The Total Variational Distance between our protocols and the quantum behaviours. \textbf{(c)} The minimum sample size needed to have at least 95\% confidence in distinguishing the two behaviours as described in the hypothesis testing scenario. In all three, the violin shapes illustrate the distributions of the values over the different projective measurements on the two-qubit state.}
    \label{fig:error}
\end{figure}

The sample size $n$ needed to distinguish the probability distributions differs with the measurement settings, with some measurement settings being more difficult to distinguish. The performance of our LHV+1 models (both the machine learning and our semianalytical approximations) over the measurement settings are given in Fig. \ref{fig:error}. It can be seen that from the neural network's protocols to our semianalytical approximations, we have gained about two orders of magnitude in Kullback-Leibler divergence. This is due to the limitations of our numerical methods used to obtain the optimum parameters, and the fact that we were bound to have missed some details from the behaviour of the network when we translated it into analytical expressions.

Our semianalytical protocols require, on average, hundreds of measurements before they can be distinguished from real quantum behaviours, disregarding other noises present in an actual quantum system. Even better, when considering the neural network themselves, it would take upwards of $10^4$ samples to distinguish them from an actual quantum system.

\final{As previously mentioned,} ideally one might try to see whether the semianalytical protocol, when integrated analytically to give the full behaviour, can be made into an exact protocol with the correct parameters. However, as the communication function is very tricky to analytically integrate, this approach might not work. On the other hand, considering that an exact protocol can already simulate some two-qubit states, these pieces of evidence suggest that all two-qubit states can be simulated with just a single bit of communication. However, ultimately, the question of \textit{exactly} simulating partially entangled states with one bit of communication remains open.

\section{Searching for Bell violation of the one-bit of communication polytope}
\label{sec:polytope}

Since two-qubit states are simulatable up to a very good precision, we now consider a different question: can we find an explicit quantum behaviour that is unsimulatable with one bit of communication? We try to go to higher dimensional systems and try to find a Bell-like inequality for the communication polytope. As far as we know, no violation of a Bell-like inequality for the one-bit of communication polytope has ever been described. \revtrois{For the rest of the section, let $\mathcal{L}$ be the local set, $\mathcal{Q}$ be the quantum set, and $\mathcal{C}$ be the one-bit of communication set. In other words, we are interested in points inside of $\mathcal{Q}$ that lie outside $\mathcal{C}$. Let also $\mathcal{A} \; (\mathcal{B})$ be the output set of Alice (Bob) and $\mathcal{X} \; (\mathcal{Y})$ her (his) input set. We will be describing scenarios with the $(|\mathcal{X}|,|\mathcal{Y}|,|\mathcal{A}|,|\mathcal{B}|)$ notation.}

\revtrois{Similar to $\mathcal{L}$, $\mathcal{C}$ is also a convex polytope. However, unlike it, it does not lie inside the no-signalling $\mathcal{NS}$ space. Bacon and Toner described the complete one bit polytope for the (2,2,2,2) scenario. They also considered the (3,3,2,2) scenario, but only for joint observables \cite{bacon2003bell}. Maxwell and Chitambar expanded these results to the (3,2,2,2) scenario later on \cite{maxwell2014bell}. Finally, the latest results are given by Cruzeiro and Gisin, which characterised the (3,3,2,2) communication polytope \cite{zambrini2019bell}. In all these works, no violation of $\mathcal{C}$ by a quantum correlation is found. Therefore, we need to go to higher dimensions. More specifically, we consider scenario with 3 outputs and beyond where Alice has more than 2 inputs (when Alice has only two inputs, she can trivially send her input to Bob and every $\mathcal{NS}$ behaviour can be simulated).}

\subsection{Description of the polytope}

\revtrois{The number of extremal points of $\mathcal{C}$ is far larger than that of $\mathcal{L}$. Indeed, the number of local deterministic strategies is $|\mathcal{A}|^{|\mathcal{X}|}|\mathcal{B}|^{|\mathcal{Y}|}$}. The number of deterministic strategies that can be performed with a single bit is $|\mathcal{A}|^{|\mathcal{X}|}|\mathcal{B}|^{2|\mathcal{Y}|}2^{|\mathcal{X}|}$ by counting; by removing duplicates, it reduces to \cite{zambrini2019bell}
$$
|\mathcal{A}|^{|\mathcal{X}|}\left(
|\mathcal{B}|^{|\mathcal{Y}|} + (2^{|\mathcal{X}|-1}-1)(|\mathcal{B}|^{2|\mathcal{Y}|}-|\mathcal{B}|^{|\mathcal{Y}|})
\right).
$$
$\mathcal{C}$ is the convex polytope formed by these vertices. In practice, we can only generate polytopes of up to around $2\times10^7$ points due to memory limitations. Since the number of extremum points for $\mathcal{C}$ is much larger than for $\mathcal{L}$, we can quickly discard the possibility of performing full facet enumeration. Hence, we would have to resort to other methods for our search.

\subsection{Random sampling quantum behaviours in higher dimensions}

We first tried to sample points from $\mathcal{Q}$ by measuring the maximally entangled two-qutrit and two-ququart state with measurements sampled uniformly in the Haar measure, before using linear programming to solve the membership problem for $\mathcal{C}$. However, this method proved ineffective as we did not manage to find any behaviour which lies \revdeux{outside} $\mathcal{C}$, and even a significant amount still lies inside $\mathcal{L}$. The statistics of this method can be seen in Table \ref{tab:montecarlo}.

\begin{table}[h!]
    \centering
    \begin{tabular}{c|cc}
    $(|\mathcal{X}|,|\mathcal{Y}|,|\mathcal{A}|,|\mathcal{B}|)$ & Points sampled & Proportion in $\mathcal{L}$ \\ \hline \hline
    (3,3,3,3) & 10000          & 25.6\% \\
    (3,4,3,3) & 300            & 10.0\% \\
    (4,3,3,3) & 300            & 10.3\% \\
    (4,4,3,3) & 100            & 1.0\% \\
    (3,3,4,4) & 500            & 56.6\% \\ \hline
    \end{tabular}   
    \caption{The statistics for the random sampling approach. None of the points sampled fall outside $\mathcal{C}$.}
    \label{tab:montecarlo}
\end{table}

\subsection{Using non-signalling points}

\begin{table*}[t]
    \centering
    \vspace{10pt}
\begin{tabular}{cc|cccc|cccc|}
&        & \multicolumn{4}{c|}{$Y=1$}          & \multicolumn{4}{c|}{$Y=2$} \\ \cline{3-10} 
&        & $B=1$ & $B=2$ & $B=3$ & $B=4$ &$B=1$ & $B=2$ & $B=3$ & $B=4$ \\ \hline
\multicolumn{1}{c|}{\multirow{4}{*}{$X=1$}} & $A=1$ & 1      & 0      & 0      & 0      & 1    & 0    & 0    & 0   \\
\multicolumn{1}{c|}{}                     & $A=2$ & 0      & 1      & 0      & 0      & 0    & 1    & 0    & 0   \\
\multicolumn{1}{c|}{}                     & $A=3$ & 0      & 0      & 1      & 0      & 0    & 0    & 1    & 0   \\
\multicolumn{1}{c|}{}                     & $A=4$ & 0      & 0      & 0      & 1      & 0    & 0    & 0    & 1   \\ \hline
\multicolumn{1}{c|}{\multirow{4}{*}{$X=2$}} & $A=1$ & 1      & 0      & 0      & 0      & 0    & 0    & 1    & 0   \\
\multicolumn{1}{c|}{}                     &  $A=2$ & 0      & 1      & 0      & 0      & 0    & 0    & 0    & 1   \\
\multicolumn{1}{c|}{}                     & $A=3$ & 0      & 0      & 1      & 0      & 1    & 0    & 0    & 0   \\
\multicolumn{1}{c|}{}                     & $A=4$ & 0      & 0      & 0      & 1      & 0    & 1    & 0    & 0   \\ \hline
\multicolumn{1}{c|}{\multirow{4}{*}{$X=3$}} & $A=1$ & 1      & 0      & 0      & 0      & 0    & 0    & 0    & 1   \\
\multicolumn{1}{c|}{}                     & $A=2$ & 0      & 1      & 0      & 0      & 0    & 0    & 1    & 0   \\
\multicolumn{1}{c|}{}                     & $A=3$ & 0      & 0      & 1      & 0      & 0    & 1    & 0    & 0   \\
\multicolumn{1}{c|}{}                     & $A=4$ & 0      & 0      & 0      & 1      & 1    & 0    & 0    & 0   \\ \hline
\multicolumn{1}{c|}{\multirow{4}{*}{$X=4$}} & $A=1$ & 1      & 0      & 0      & 0      & 0    & 1    & 0    & 0   \\
\multicolumn{1}{c|}{}                     & $A=2$ & 0      & 1      & 0      & 0      & 1    & 0    & 0    & 0   \\
\multicolumn{1}{c|}{}                     & $A=3$ & 0      & 0      & 1      & 0      & 0    & 0    & 0    & 1   \\
\multicolumn{1}{c|}{}                     & $A=4$ & 0      & 0      & 0      & 1      & 0    & 0    & 1    & 0   \\ \hline
\end{tabular}
    \caption{\final{The correlation table for} the point in $(4,2,4,4)$ which have $w_\mathcal{Q}=w_\mathcal{C}=\frac{2}{3}$. Each of the smaller eight boxes corresponds to the output table for a particular combination of inputs, which Alice's input indexing the vertical dimension and Bob's the horizontal. In each of the boxes, the $4 \times 4$ table corresponds to the outputs of Alice and Bob, with Alice's in the vertical and Bob's in the horizontal. Note that the 1 here means $\frac{1}{4}$, which means that each of the boxes sum up to 1, as required.}
    \label{tab:4244}
\end{table*}

\begin{figure}[h!]
    \centering
    \input{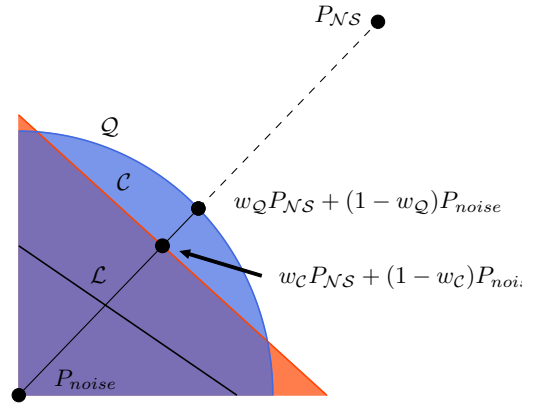}
    \caption{$w_\mathcal{Q}$ is the threshold \final{correlation} weight for the quantum set $\mathcal{Q}$, while $w_\mathcal{C}$ is the threshold \final{correlation} weight for the one-bit communication set $\mathcal{C}$. Thus, \revdeux{$w_\mathcal{C}<w_\mathcal{Q}$} would imply a violation and would give a quantum behaviour that could not be simulated by a single bit of communication.}
    \label{fig:diagram}
\end{figure}

The next method we used was to use points in $\mathcal{NS}$ ($P_{\mathcal{NS}}$) and mix them with noise ($P_{noise}$) to obtain
$$
wP_{\mathcal{NS}} + (1-w)P_{noise}
$$
for some correlation weight $0 \leq w \leq 1$. We do this in order to find out the threshold \final{weights} at which the correlation exits the sets $\mathcal{Q}$ and $\mathcal{C}$. If we find a correlation $P_{\mathcal{NS}}$ which exits $\mathcal{Q}$ at a \final{higher correlation weight} $w_\mathcal{Q}$ than the corresponding one for $\mathcal{C}$, $w_\mathcal{C}$, all behaviours with $w_\mathcal{C}<w<w_\mathcal{Q}$ would be behaviours in $\mathcal{Q}$ that are unsimulatable by one-bit of communication. Thus, we will focus on finding the smallest value of the gap ($w_\mathcal{C} - w_\mathcal{Q}$), where a negative value of the gap would signify a violation. A graphical illustration can be seen in Fig. \ref{fig:diagram}.

The membership problem for $\mathcal{Q}$ is solved using the NPA hierarchy method \cite{navascues2008convergent} with level 2 hierarchy\final{, which we ran using QETLAB \cite{johnston2016qetlab}}. \revtrois{Note that the NPA hierarchy gives only an upper bound on $\mathcal{Q}$.} As before, the membership problem for $\mathcal{C}$ is solved using linear programming.

Unfortunately, choosing a suitable $P_\mathcal{NS}$ proved to be a challenge. The extremum points of the $\mathcal{NS}$ space have only been characterised for binary inputs or binary outputs \cite{barrett2005nonlocal, jones2005interconversion}. Here, we mostly used nonlocal points which are locally unbiased, i.e. for all inputs, all the local outputs are of equal probability, and maximally correlated, i.e. for all input combinations, there is a perfect correlation between Alice and Bob's outputs, for a particular output of Alice the output of Bob is guaranteed and vice versa. While we tried other non-signalling points, this particular class of points gave us the smallest gap ($w_\mathcal{C} - w_\mathcal{Q}$) in all scenarios. Similarly, there are also numerous choices for $P_{noise}$, but we found that white noise gives the smallest gap in most scenarios.

\revdeux{The results of the smallest gap ($w_\mathcal{C} - w_\mathcal{Q}$) found in each scenario we investigated are listed in Table \ref{tab:NS_table}. In the case of $|\mathcal{A}|=|\mathcal{B}|=3$, we did not find any violation. The smallest gap was observed in the $(|\mathcal{X}|,|\mathcal{Y}|,|\mathcal{A}|,|\mathcal{B}|)=(4,4,3,3)$ scenario, where a there exist a point with $w_\mathcal{C}=0.6612$ and $w_\mathcal{Q}=0.6289$.} However, in $|\mathcal{A}|=|\mathcal{B}|=4$, specifically in the $(4,2,4,4)$ setting, we found a $P_\mathcal{NS}$ which have $w_\mathcal{Q}=w_\mathcal{C}=\frac{2}{3}$, described in Table \ref{tab:4244}. The table itself can also be interpreted as a Bell inequality by taking the terms in the table as the coefficients of the correlation terms and adding all of them. When normalised into a Bell game, the value of the game is both $\frac{3}{4}$ for $\mathcal{Q}$ and $\mathcal{C}$. \revother{Geometrically, this Bell inequality} is the hyperplane which has the line connecting $P_{\mathcal{NS}}$ and $P_{noise}$ as its normal.

\revtrois{The maximum quantum bound of this inequality can be obtained by the maximally entangled two-ququart state,
$$
\frac{1}{4} (\ket{00} + \ket{11} + \ket{22} + \ket{33}),
$$
with suitable measurements found with heuristic numerical methods. Equivalently, with the same measurements, this state is able to realise the correlation
$
w_{\mathcal{Q}}P_{\mathcal{NS}} + (1-w_{\mathcal{Q}})P_{noise},
$
where $P_{\mathcal{NS}}$ is the correlation in Table \ref{tab:4244} and the $w_{\mathcal{Q}} = \frac{2}{3}$ as previously mentioned. This explicit behaviour meets the NPA bound to numerical precision, and thus we can conclude that the maximum correlation weight $w_{\mathcal{Q}}=\frac{2}{3}$ and the quantum bound of $\frac{3}{4}$ for the inequality are exact in this case.}

This point represents our closest attempt at finding a violation with this method. The number of extremal points for $\mathcal{C}$ in the $(4,2,4,4)$ scenario is around $1\times10^6$, and it is still possible to go one input higher to $(4,3,4,4)$ or $(5,2,4,4)$. A violation might exist there, but our heuristic search proved unfruitful. In the end, contrary to the prepare-and-measure scenario \cite{renner2023classical}, it still remains an open problem to find a bipartite quantum behaviour that is provably unsimulatable with one bit of communication \footnote{One might hope that the prepare-and-measure (P\&M) example \cite{renner2023classical} could be an entry point; but it turns out not to be. Indeed, consider the corresponding entanglement-based scenario, in which one steers probabilistically the P\&M behaviors $P(b|u=(x,a),y)$ from some bipartite correlations $P(a,b|x,y)$. However, since the P\&M examples only use one qubit and a projective measurement by Bob, the bipartite behavior can be simulated with at most one trit  \cite{renner2022minimal}, and probably one bit as the first part of our paper proves.}.

\begin{table}[]
    \centering
    \begin{tabular}{c|cc|c}
    $(|\mathcal{X}|,|\mathcal{Y}|,|\mathcal{A}|,|\mathcal{B}|)$     & $w_\mathcal{Q}$ & $w_\mathcal{C}$ & \revdeux{$w_\mathcal{C} - w_\mathcal{Q}$} \\
    \hline
    \hline
    (3,3,3,3) & 0.5995 & 0.7000 & 0.1005\\
    (3,4,3,3) & 0.6022 & 0.7000 & 0.0978\\
    (4,3,3,3) & 0.5856 & 0.6766 & 0.0910 \\
    (4,4,3,3) & 0.6289 & 0.6612 & 0.0323\\
    (5,3,3,3) & 0.5768 & 0.6610 & 0.0842\\
    (3,3,4,4) & 0.6159 & 0.7143 & 0.0984\\
    (4,2,4,4) & 0.6666 & 0.6666 & 0.0000\\
    \hline
    \end{tabular}
    \caption{\revdeux{The smallest gap ($w_\mathcal{C} - w_\mathcal{Q}$) in each scenario we studied.}}
    \label{tab:NS_table}
\end{table}

\section{Conclusion}

In this work, we tried to further the works that have been done on characterising the communication complexity cost of quantum behaviours. We tried to obtain a protocol to simulate partially entangled two-qubit states using a neural network, and we presented a semianalytical LHV+1 protocol based on the protocol of the neural networks. While these protocols could only approximate the quantum behaviours, on average one needs hundreds of measurement data, for the semianalytical protocols, and tens of thousands of measurement data, for the neural network protocols, in order to be distinguished from the quantum behaviour. We also tried to find quantum behaviours in higher dimensions that could not be simulated with one bit of communication. While we were able to find a Bell-like inequality that has the same maximum value in $\mathcal{Q}$ and $\mathcal{C}$, we were unable to find a violation.

From this work and all the previous works done on the topic, it can be seen that evaluating the capabilities of entangled quantum states in terms of communication complexity is very difficult. While we are confident that a behaviour that cannot be simulated with a single bit could probably be found, extending the work to more bits and states would probably be too difficult, barring any new revolutionary techniques. On the other hand, from our result that numerical protocols that closely approximate the two-qubit entangled states can be found, the task of simulating partially entangled two-qubit states using one bit of communication \textit{exactly} is probably possible and a fully analytical protocol could probably be found in the near future.

\revtrois{\textit{Note added in proof}: while the review of this paper was being finalised, Márton et al.~showed how to find quantum correlations outside $\mathcal{C}$ \cite{marton2023beating} by using parallel repetition. The Bell scenarios, for which such examples have been found, have many more inputs and/or outputs than those studied here.}

\section*{Code Availability}
The code is available at \hyperlink{https://github.com/PeterSidajaya/neural-network-fyp/}{https://github.com/PeterSidajaya/neural-network-fyp/}.

\section*{Acknowledgments}
This research is supported by the National Research Foundation, Singapore and A*STAR under its CQT Bridging Grant. We thank Maria Balanz\'o-Juand\'o, Nicolas Gisin, Marco T\'ulio Quintino, Martin J. Renner, and Marco Tomamichel for discussions. We are also grateful to the authors of \cite{krivachy2020neural} for making their code public.

We also thank the National University of Singapore Information Technology for the use of their high performance computing resources.

\bibliographystyle{unsrtnat}
\bibliography{ref}

\onecolumn
\appendix

\section{Coefficients for semianalytical models}
\label{app:tables}

The parameters listed in \final{Table \ref{tab:party_coefficients_1}} are obtained using an evolutionary algorithm. The algorithm works by first randomly generating a population of numbers as candidates for the parameters. Then, the fitnesses of the candidates are evaluated by comparing them to the target, here being the neural network's outputs. Finally, only the top candidates are kept, and the population is replenished by crossover and mutations amongst the remaining candidates.

\begin{table}[h!]
    \centering
    \begin{tabular}{c|c|cc|ccc}
        \hline \hline
        $\alpha$ & Party & u & v & w & x & y \\ \hline \hline
        \multirow{5}{*}{$\frac{\pi}{4}$}
            & Alice 1 & -0.9121 & 0.2046 & -0.0052 & 0.0065 & 0.0041 \\
            & Alice 2 & 0.9143 & 0.1907 & -0.0011 & -0.0048 & -0.0170 \\
            & Bob 1 & -0.9230 & 0.1832 & 0.0067 & 0.0030 & -0.0060 \\
            & Bob 2 & 0.9645 & 0.2044 & -0.0022 & -0.0005 & 0.0155 \\
            & Comm & -0.0082 & 0.0257 &&& \\ \hline \hline
        \multirow{5}{*}{$\frac{7\pi}{32}$}
            & Alice 1 & -0.7729 & 0.0699 & 0.0022 & -0.0039 & 0.0045 \\
            & Alice 2 & 0.8185 & 0.2886 & -0.0110 & 0.0004 & -0.0003 \\
            & Bob 1 & -0.8280 & 0.0762 & 0.0035 & -0.0066 & -0.0059 \\
            & Bob 2 & 0.7510 & 0.2746 & -0.0004 & 0.006 & 0.0037 \\
            & Comm & 0.0019 & 0.1640 &&& \\ \hline \hline
        \multirow{5}{*}{$\frac{3\pi}{16}$}
            & Alice 1 & -0.7198 & -0.0008 & -0.0054 & -0.0180 & 0.0396 \\
            & Alice 2 & 0.7454 & 0.5065 & -0.0040 & -0.0025 & -0.0138 \\
            & Bob 1 & -0.7809 & 0.0420 & 0.0302 &  0.0177 & -0.0346 \\
            & Bob 2 & 0.6458 & 0.4146 & 0.0318, & 0.0835 & 0.0536 \\
            & Comm & 0.0718 & 0.3152 &&& \\ \hline \hline
        \multirow{5}{*}{$\frac{5\pi}{32}$}
            & Alice 1 & -0.6575 & -0.0320 & 0.0645 & 0.2220 & -0.2995 \\
            & Alice 2 & 0.6348 & 0.7453 & -0.0526 & -0.2358 & -0.0718 \\
            & Bob 1 & -0.7236 & 0.0786 & -0.0013 & -0.0127 & -0.0176 \\
            & Bob 2 & 0.4924 & 0.6380 & -0.0060 & 0.0032 & -0.0130 \\
            & Comm & 0.1313 & 0.3838 &&& \\ \hline \hline
        \multirow{5}{*}{$\frac{\pi}{8}$}
            & Alice 1 & -0.7401 & 0.0687 & -0.0567 & -0.2704 & 0.2295 \\
            & Alice 2 & 0.2644 & 1.1938 & -0.1204 & -0.6588 & -0.0845 \\
            & Bob 1 & -0.7541 & 0.0913 & 0.0134 & 0.0116 & 0.0029 \\
            & Bob 2 & 0.2903 & 1.0314 & -0.0047 & 0.0302 & 0.0002 \\
            & Comm & 0.1768 & 0.2272 &&& \\ \hline \hline
    \end{tabular}
    \caption{The coefficients for the network's outputs for different $\alpha$.}
    \label{tab:party_coefficients_1}
\end{table}

To get an estimate of the errors in these parameter values, one can look at the values for $\alpha = \frac{\pi}{4}$ and compare them with the maximally entangled state protocol. There, the values for $w,x,y$ are all supposed to be 0 for the exact analytical protocol. Moreover, for the parties, \revtrois{$|u|=1$ and $v=0$ should hold}. Finally, $u$ and $v$ for the communication protocol should also be 0. From there, it can be inferred that the values of $u$ and $v$ in our semianalytical protocol are quite distant from the optimal value of the parameter. This is one of the reasons for the two orders of magnitude of error obtained when translating the behaviour of the neural network into a semianalytical protocol.

\section{Figures}\label{app:figures}

In the following figures, we will show the output of the neural networks in terms of figures similar to the ones found in \cite{toner2003communication}, \revdeux{but in polar coordinates (in $\theta$ and $\phi$). These figures show how Alice and Bob would respond in a given round for a certain fixed pair of LHV. I.e., the two "Alice" plots are the plots of the function
$$P(A_i=+1 \mid \Vec{a}, \lambda = \lambda_{fixed})$$
as a function of $\Vec{a}$ for $i\in\{1,2\}$. Similarly for the "Communication" plot ($P(c=+1 \mid \Vec{a}, \lambda = \lambda_{fixed})$) and the "Bob" plots ($P(B_i=+1 \mid \Vec{b}, \lambda = \lambda_{fixed})$). Thus, the axes are the input of Alice or Bob, depending on the plots. The pink and cyan dots are the position of the LHV, which are uniformly distributed throughout the spheres, but are fixed in these figures to show the shape of the functions.}

\begin{figure*}[h!]
    \centering
    \includegraphics[width=0.3\textwidth]{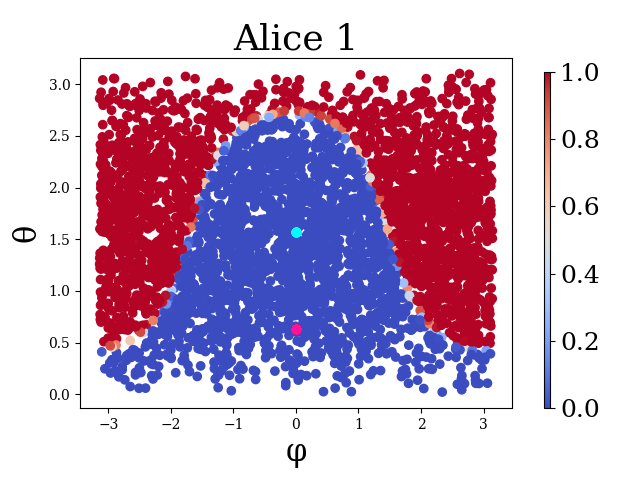}
    \includegraphics[width=0.3\textwidth]{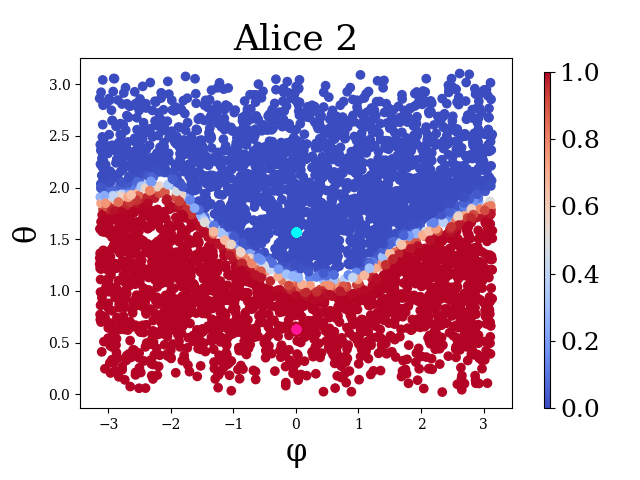}
    \includegraphics[width=0.3\textwidth]{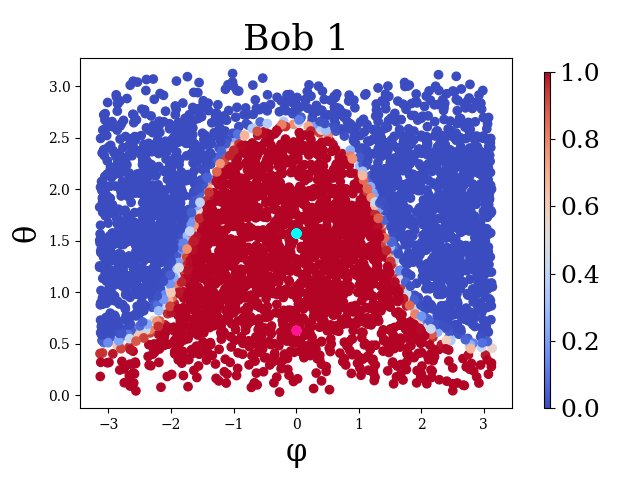}
    \includegraphics[width=0.3\textwidth]{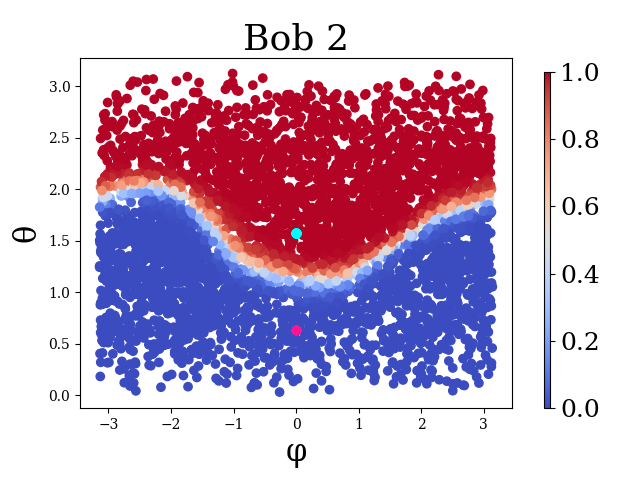}
    \includegraphics[width=0.3\textwidth]{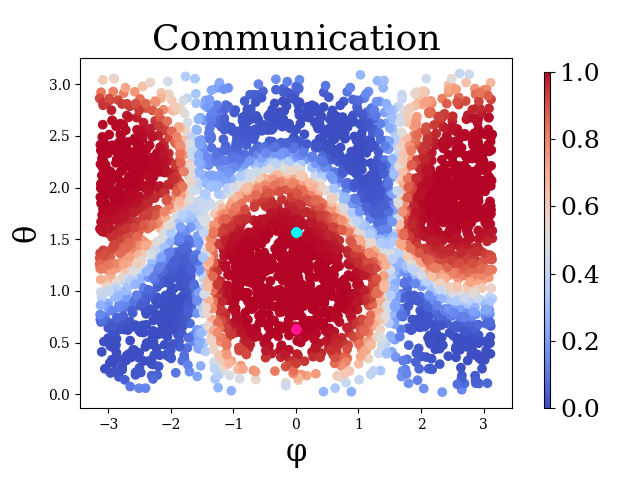}
    \caption{The protocol generated by the ANN for $\ket{\psi(\frac{\pi}{4})}$. \revdeux{Here the axes are the Bloch sphere in polar coordinates. In the "Alice" and "Communication" plots, they show the direction of $\Vec{a}$. In the "Bob" plots, they show the direction of $\Vec{b}$. The colour is the output of the $A$, $B$, $c$ functions of the neural network. We give an analytical expression of these functions in Section \ref{subsubsec:max-entangled}.} While the lines are not smooth, the protocol suggests more or less a Toner-Bacon like protocol.}
    \label{fig:pi-4}
\end{figure*}

\begin{figure*}[h!]
    \centering
    \includegraphics[width=0.3\textwidth]{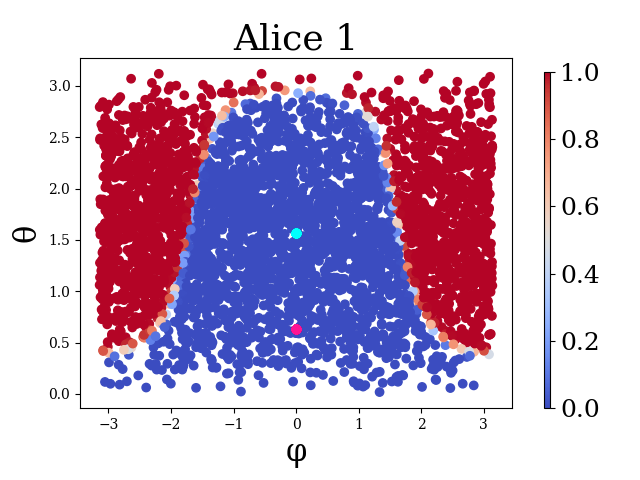}
    \includegraphics[width=0.3\textwidth]{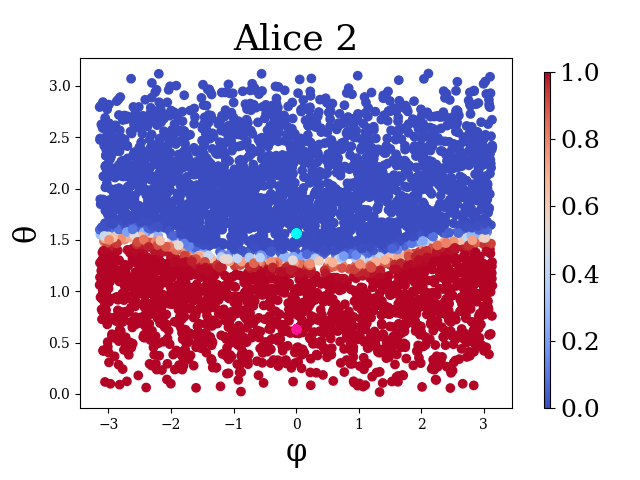}
    \includegraphics[width=0.3\textwidth]{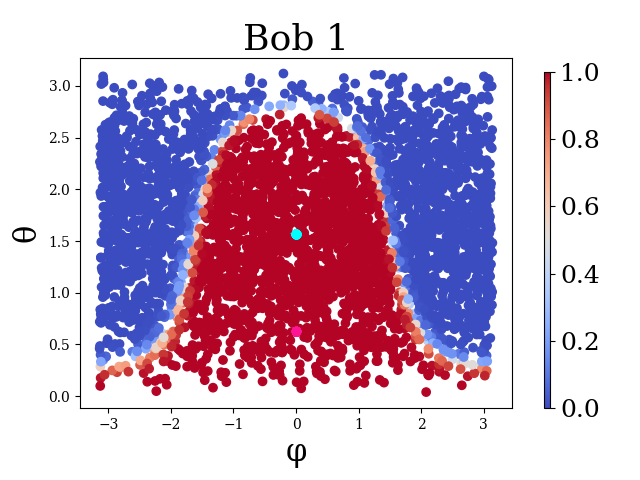}
    \includegraphics[width=0.3\textwidth]{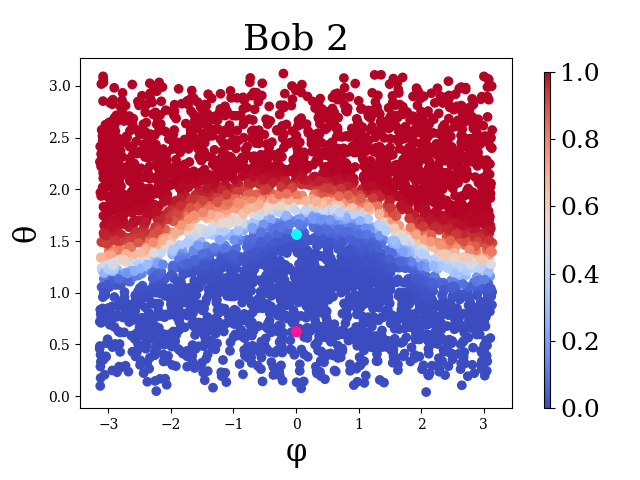}
    \includegraphics[width=0.3\textwidth]{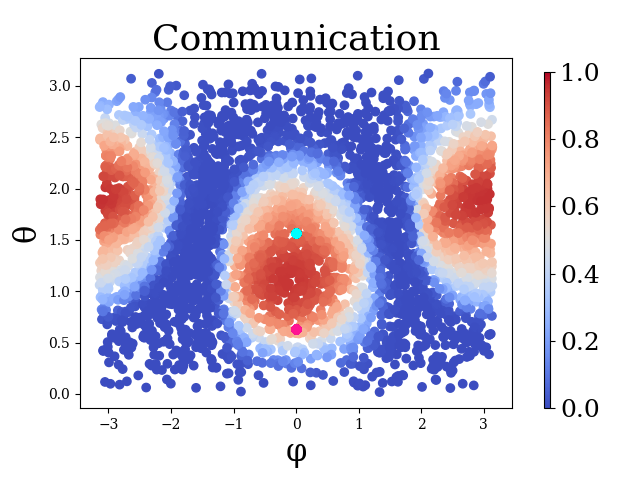}
    \caption{The protocol generated by the ANN for $\ket{\psi(\frac{5\pi}{32})}$. \revother{Notice the that now the transitions between the red and blue regions are less sharp than the one of the maximally entangled state. This means that now the bit of communication sent is nondeterministic in those areas}. This model is trained starting from the $\ket{\psi(\frac{\pi}{4})}$ model shown above, thus it has similar features.}
    \label{fig:5pi-32} 
\end{figure*}

\begin{figure*}[h!]
    \centering
    \includegraphics[width=0.3\textwidth]{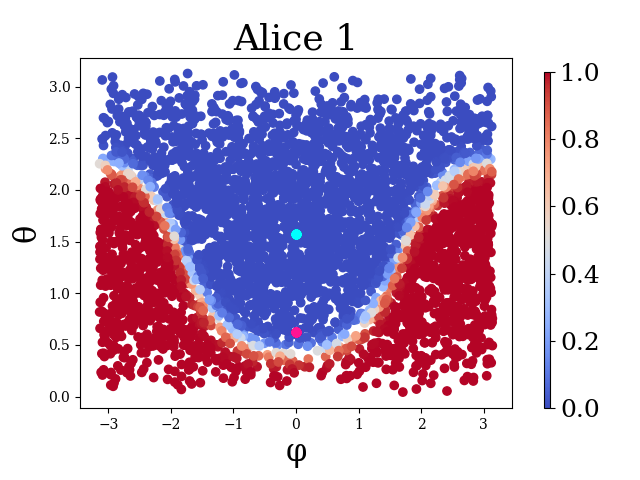}
    \includegraphics[width=0.3\textwidth]{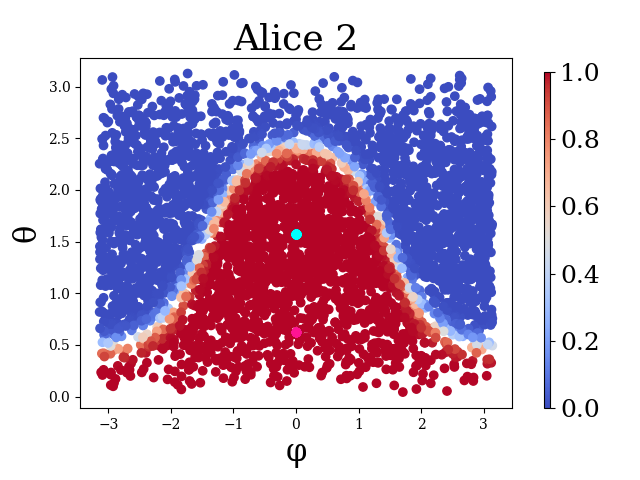}
    \includegraphics[width=0.3\textwidth]{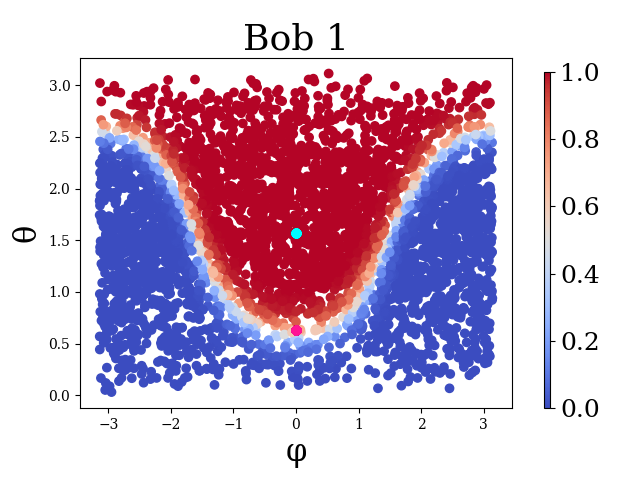}
    \includegraphics[width=0.3\textwidth]{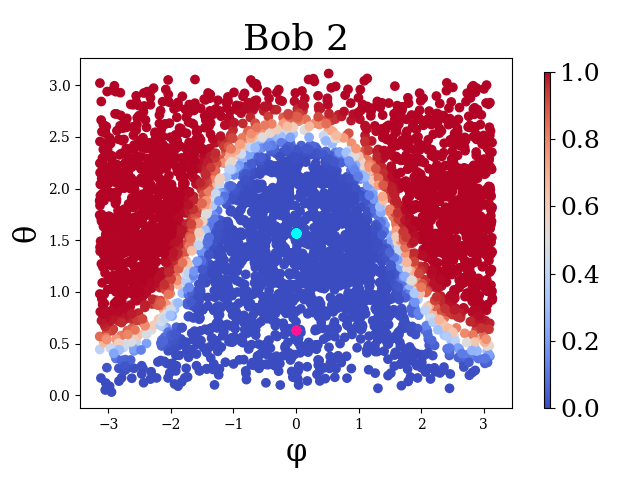}
    \includegraphics[width=0.3\textwidth]{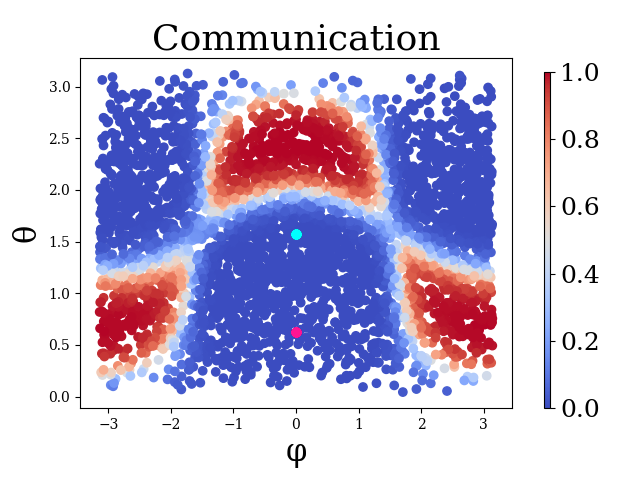}
    \caption{The protocol generated by the ANN for $\ket{\psi(\frac{5\pi}{32})}$, but with the caveat that the bit of communication is now of a simpler form. Notice that now, the bit of communication sent is deterministic again. The forms of the parties' outputs are slightly different from the previous figures (Fig. \ref{fig:5pi-32}) because they are trained from different starting models.}
    \label{fig:5pi-32-simple} 
\end{figure*}

\end{document}